\documentclass{article}
\usepackage{graphicx}
\usepackage{url}

\usepackage{amssymb}
\usepackage{amsmath}
\usepackage{latexsym}

\newtheorem{theorem}{Theorem}

\title{Computational Geometry Column 43}
\author{%
Joseph O'Rourke\thanks{
Dept. of Computer Science, Smith Col\-lege, North\-ampton, 
MA 01063, USA.
orourke@\allowbreak cs.\allowbreak smith.\allowbreak edu.
Supported by NSF Distinguished Teaching Scholar Grant DUE-0123154.
}
}
\date{}
\begin{document}
\maketitle

\begin{abstract}
The concept of pointed pseudo-triangulations is defined
and a few of its applications described.
\end{abstract}

A \emph{pseudo-triangle} is a planar polygon with exactly
three convex vertices.  Each pair of convex vertices is connected
by a reflex chain, which may be just one segment.  Thus, a triangle
is a pseudo-triangle.
A \emph{pseudo-triangulation} of a set $S$ of $n$ points in the plane
is a partition of the convex hull of $S$ into pseudo-triangles
using $S$ as a vertex set.
Under the name of ``geodesic triangulations,'' these found
use in ray shooting~\cite{cegghss-rspug-94},
and then for visibility algorithms by 
Pocchiola and Vegter~\cite{pv-cvgpt-95},
who named them (after a dual relationship to pseudoline arrangements)
and studied many of their properties~\cite{pv-ptta-96}.
Recently the identification of \emph{minimum pseudo-triangulations}
by Streinu~\cite{s-capnc-00} has generated a flurry of
new applications that we selectively sample here.

\begin{figure}[htbp]
\centering
\includegraphics[width=0.7\linewidth]{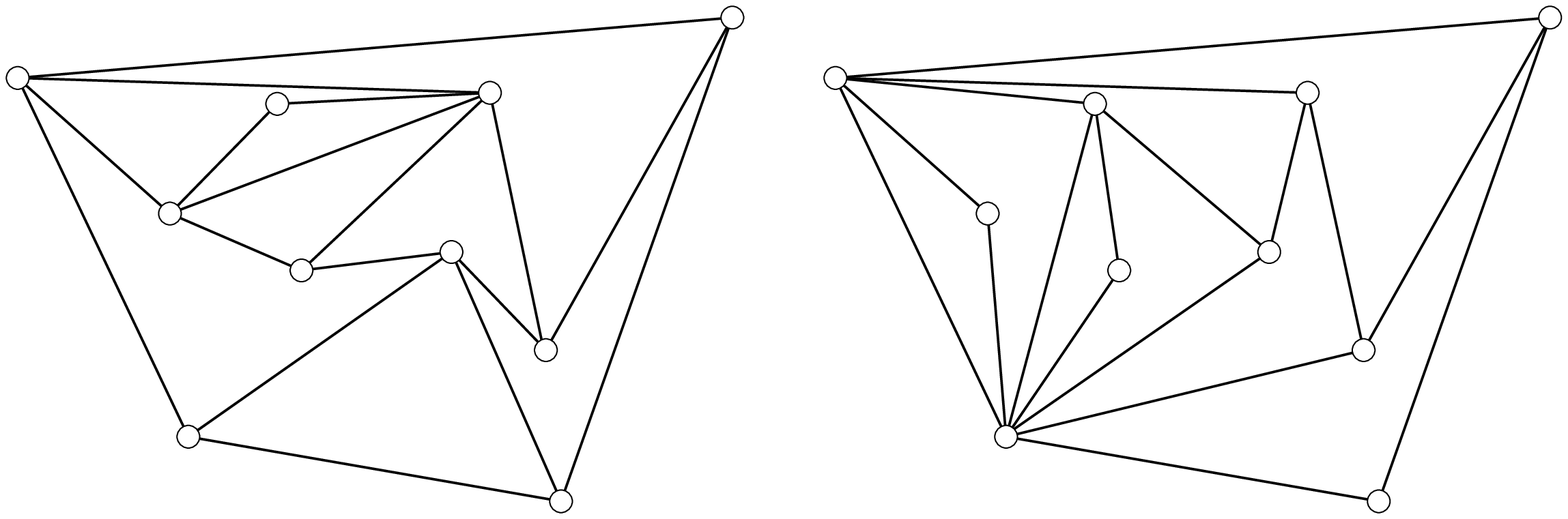}
\caption{Two pointed pseudo-triangulations of the same set of $n=10$ points.}
\label{pseudotri}
\end{figure}

Minimum pseudo-triangulations have the fewest possible number of
edges for a given set $S$ of points.
The term \emph{pointed pseudo-triangulation} is gaining prominence
because of the following theorem\footnote{
	Thm.~3.1(2) of~\cite{s-capnc-00} (where the property is 
	called ``acyclicity'');
	Lem.~1 in~\cite{kkmsst-tdbpt-01}.
}
(see Fig.~\ref{pseudotri}):
\begin{theorem}
A pseudo-triangulation of $S$ is minimum iff
every point $p \in S$ is \emph{pointed} in the sense
that its incident edges span less than $\pi$
(i.e., they fall within a cone apexed at $p$ with aperture angle $< \pi$).
\end{theorem}


Pointed pseudo-triangulations
have remarkable regularity in that each contains 
exactly $n-2$ pseudo-triangles.
In this they parallel triangulations of simple polygons and are unlike
triangulations of point sets, which may have from $n-2$ to $3(n-2)$ triangles.
Another indication of their well-behavedness
is the recent result~\cite{kkmsst-tdbpt-01}
that every set of $n$ points in general position
has a pointed pseudo-triangulation with maximum vertex degree
of $5$, a property not enjoyed by triangulations
(of either polygons or point sets).
This bounded degree may prove useful for designing data structures
employing pseudo-triangulations, such as those used for
collision detection~\cite{kss-kcdsp-02}.

We mention two disparate applications of pointed pseudo-triangulations.
Two years ago a long-standing open problem was settled
by Connelly, Demaine, and Rote when they proved that 
polygonal linkages cannot lock in the plane:
open
polygonal chains can be straightened, and closed polygonal
chains can be convexified~\cite{cdr-spacp-00}~\cite{o-cgc39-00}.
Their unlocking motions are \emph{expansive}: no distance
between any pair of vertices decreases.
Their proof relies on proving the existence of a solution to
differential equations describing this motion.
Although not difficult to solve numerically in practice,\footnote{
        \url{http://db.uwaterloo.ca/~eddemain/linkage/animations/}.
}
theirs does not constitute an algebraic proof.
This was provided by Streinu~\cite{s-capnc-00}.
She proved that a pointed pseudo-triangulation is rigid
when viewed as a bar and joint framework, and that
removal of a convex hull edge yields a mechanism with
one degree of freedom (1-dof) that executes an expansive motion.
Adding bars to a given collection of polygonal chains
to produce a pointed pseudo-triangulation,
and using a useful ``flipping'' property of pseudo-triangulations, leads 
ultimately to
a piecewise-algebraic straightening/convexifying motion.
These 1-dof expansive motions have a rich structure that is just now
being elucidated~\cite{rss-emppp-01}.

Second, pointed pseudo-triangulations were used to make an
advance on an open problem posed by Urrutia~\cite{u-agip-00}:
Can a simple polygon of $n$ vertices be illuminated by
$c n$ interior $\pi$-floodlights placed at vertices,
with $c < 1$ (for sufficiently large $n$)?
For several years the lower bound on $c$ was $3/5$, but
no one could prove that $c$ was smaller than $1$.
Starting from a pointed pseudo-triangulation of the polygon,
Speckmann and T\'{o}th have recently established~\cite{st-vpgsp-01} that
$(2/3)n$ lights suffice,
finally breaking the $c=1$ barrier.\footnote{
        See \url{http://cs.smith.edu/~orourke/TOPP/}, Problem 23,
        for further developments.
}

We close with an open problem, posed in~\cite{rrss-ctpw-01}.
For a given set $S$ of points, let $\#T$ be the number
of triangulations of $S$, and $\#PPT$ be the number
of pointed pseudo-triangulations of $S$.
They conjecture that $\#T \le \#PPT$ for all $S$,
with equality only when the points of $S$ are in convex position.
This has been established for all sets of at most $9$ points~\cite{bkps-ceptg-01}.

\small
\bibliographystyle{alpha}
\bibliography{43,/home1/orourke/bib/geom/geom}

\end{document}